\documentclass{optica-article}

\journal{opticajournal} % for journals or Optica Open

\articletype{Research Article}

\usepackage{lineno}
\usepackage{float}
\usepackage{graphicx}
\usepackage{siunitx}
\usepackage[table]{xcolor}
\newcommand\hlr[1]{%
  \bgroup
  \hskip0pt\color{red!80!black}%
  #1%
  
  \egroup
}

\begin{document}

\title{Optically-triggered deterministic spiking regimes in nanostructure resonant tunnelling diode-photodetectors}

\author{Qusay Raghib Ali Al-Taai\authormark{1}, 
Matěj Hejda\authormark{2}, 
Weikang Zhang\authormark{2},
Bruno Romeira\authormark{3},
José M. L. Figueiredo\authormark{4},
Edward Wasige\authormark{1},
Antonio Hurtado\authormark{2}*}

\address{\authormark{1}High Frequency Electronics Group, University of Glasgow, Glasgow, United Kingdom\\
\authormark{2}Institute of Photonics, SUPA Dept of Physics, University of Strathclyde, Glasgow, United Kingdom\\
\authormark{3}INL – International Iberian Nanotechnology Laboratory, Ultrafast Bio- and Nanophotonics Group, Braga, Portugal\\
\authormark{4}Centra-Ci\^{e}ncias and Departamento de F\'{i}sica, Faculdade de Ci\^{e}ncias, Universidade de Lisboa, Lisboa, Portugal}

\email{\authormark{*}antonio.hurtado@strath.ac.uk} %% email address is required; see note below about the corresponding author designation

% use {asbstract*} to suppress the copyright line. Copyright information will be added in production

\begin{abstract*} 
This work reports a nanostructure resonant tunnelling diode-photodetector (RTD-PD) device and demonstrates its operation as a controllable, optically-triggered excitable spike generator. 
The top contact layer of the device is designed with a nanopillar structure (\SI{500}{\nano\metre} in diameter) to restrain the injection current, yielding therefore lower energy operation for spike generation. We demonstrate experimentally the deterministic optical triggering of controllable and repeatable neuron-like spike patterns in the nanostructure RTD-PDs. Moreover, we show the device’s ability to deliver spiking responses when biased in both regions adjacent to the negative differential conductance (NDC) region, the so-called ‘peak’ and ‘valley’ points of the current-voltage ($I$-$V$) characteristic. This work also demonstrates experimentally key neuron-like dynamical features in the nanostructure RTD-PD, such as a well-defined threshold (in input optical intensity) for spike firing, as well as the presence of spike firing refractory time. The optoelectronic and chip-scale character of the proposed system together with the deterministic, repeatable and well controllable nature of the optically-elicited spiking responses render this nanostructure RTD-PD element as a highly promising solution for high-speed, energy-efficient optoelectronic artificial spiking neurons for novel light-enabled neuromorphic computing hardware.
\end{abstract*}

%%%%%%%%%%%%%%%%%%%%%%%%%%  body  %%%%%%%%%%%%%%%%%%%%%%%%%%
% Sample is R4C9_6x6window
\section{Introduction}
The current dramatic increase in data processing demands requires new computing systems for Artificial Intelligence (AI) functionalities that enable increased processing speeds as well as optimized power consumption. However, traditional digital computing systems conventionally based on the Von Neumann architecture are not well suited for the massively-parallel character of artificial neural network (ANN) algorithms. Neuromorphic computing is among the main research avenues investigating novel computing architectures, drawing inspiration from the brain and aiming at reproducing its energy efficiency, extensive parallelism and powerful computational capabilities. The human brain has a topological structure consisting of billions of neurons connected by trillions of synapses through axons and dendrites \cite{Liang2016_NeuralInfoProc}. Neurons are the fundamental building blocks of information processing in the brain, typically using signal propagation in the form of time-dependent sparse spikes. Spiking is an event-driven computation scheme that primarily consumes energy only when information processing occurs. In particular, neuromorphic computing realized with photonics shows the potential to outperform digital electronic counterparts in computing speed, bandwidth and energy efficiency \cite{Shastri2021_NatPhot}. Among the photonic neuromorphic approaches, those that offer the functionality of neuron-like spiking and excitability are of significant interest. Excitability represents a type of behaviour observed in dynamical systems, characterized by all-or-nothing responses (typically called spikes). Excitable systems typically exhibit two essential properties: \textit{thresholding} and \textit{refractoriness}. Thresholding indicates the ability of a system (such as a biological neurons) to fire spikes only in response to super-threshold stimuli while it remains quiescent otherwise. Thresholding plays an essential role in controlling the states in neuromorphic processing elements for both the filtering and processing of incoming signals, as well as for allowing the system to decide whether to elicit a spike to be propagated to other neurons \cite{Prucnal2016_AOP}. 

Spiking has been shown in all-optical devices such as vertical cavity surface emitting lasers under injection locking \cite{Robertson2020_SciRep, Skalli2022_OME}, two-section lasers \cite{Song2020_IJSTQE,Ma2018_OL} and phase-change material (PCM) based architectures \cite{Chakraborty2018_SciRep}. Furthermore, optoelectronic approaches utilizing photodetectors for spike activation provide good degree of flexibility by combining excitable electronic devices with photonic components. Such systems have been demonstrated in superconducting Josephson junctions \cite{Shainline2018_JApplPhys}, biomimetic circuits based on multilayer MoS$_2$ 2D material with silicon photodetector \cite{Radhakrishnan2021_NatComm}, Ge/Si photonic crystal-enhanced photodiodes coupled to excitable circuitry and a quantum-dot laser \cite{Lee2022_OE}, high-speed integrated balanced photodetectors coupled to an on-chip, two-section DFB laser \cite{Peng2020_IJSTQE}, as well as optoelectronic circuits based upon resonant tunnelling diodes (RTDs) \cite{Romeira2016_SciRep}.

The RTD is an active semiconductor device typically based on an embedded, double-barrier quantum well (DBQW) heterostructure that introduces ultrafast quantum tunnelling phenomena. Thanks to the filtering effect of the DBQW on charge carriers wavefunction, the RTD exhibits unusual, N-shaped current-voltage ($I$-$V$) characteristic with a negative differential conductance (NDC) region. This highly nonlinear $I$-$V$ leads to a range of nonlinear dynamical behaviours including excitability \cite{Lourenco2022_JPCS}, which allows for generation of high speed neuron-like spiking signals \cite{Romeira2013_OE, Romeira2014_OQE}. This functionality was utilized to experimentally realize a self-feedback, neuron-like autaptic cell \cite{Romeira2016_SciRep}, and to numerically demonstrate a feedforward spiking artificial neural network numerical model powered by optoelectronic RTD nodes \cite{Hejda2022_PRAppl}. In addition, the RTD has a high optical sensitivity when being designed as a photodetector thanks to the intrinsic gain and wide frequency bandwidth operating up to terahertz (THz) range. Various modes of operation for optically modulated RTDs have been previously demonstrated for photodetection and wireless systems, allowing the measurement of light responsivity (e.g., as reviewed in \cite{Watson2019_MOTL}) with choice of a suitable epitaxial structure enhancing light sensitivity \cite{Wang2017_TICAOP}.

This work focuses on the fabrication, experimental characterisation and dynamical analysis of a nanostructure photo-detecting RTD device (RTD-PD) for use as a light-driven optoelectronic spiking neuron. The embedded nanopillar permits operation with reduced bias currents \cite{Al-Taai2021_IJNM} for more energy efficient operation, with a micrometre-scale optical window for enhanced light detection and deterministic optical triggering of electrical spiking. Furthermore, the operation can be extended from optical-to-electrical (O/E) spiking to fully opto-electro-optical O/E/O node by coupling such node to a light source, such as a laser \cite{Hejda2022_Nanophot}.

\section{Nanostructure RTD-PD: Design and Fabrication}
This section describes the design and fabrication of the nanostructure RTD-PD, with device layers being described here under the assumption of forward biasing. We will refer to two design parameters that describe the scale of RTD devices: the area of the top contact $S_{TC}$ and a total area of light absorption region $S_{total}$ (without any metal contact).

To reduce the power consumption and enable energy-efficient operation and spike generation in the RTD, a new design is presented where the top collector of the RTD-PD is shrunken down to the shape of a nanopillar (also be referred to as a nano-injector \cite{Pfenning2022_Nanomaterials}). When compared to RTD-PD devices with uniform micrometre-scaled layers, the nanopillar structure reduces the current on the collector side and reintroduces the current path into the DBQW region of the device, therefore reducing the peak current ($I_P$). The characterization of the electrical static properties of similar nanostructure RTD-PDs has been reported in our previous work \cite{Al-Taai2021_WOCDISE} showing that this new nanopillar-incorporating design introduces an important reduction in the peak current of the device, thus significantly decreasing overall power consumption.

\begin{figure}[h!]
    \centering
    \includegraphics[width=0.95\linewidth]{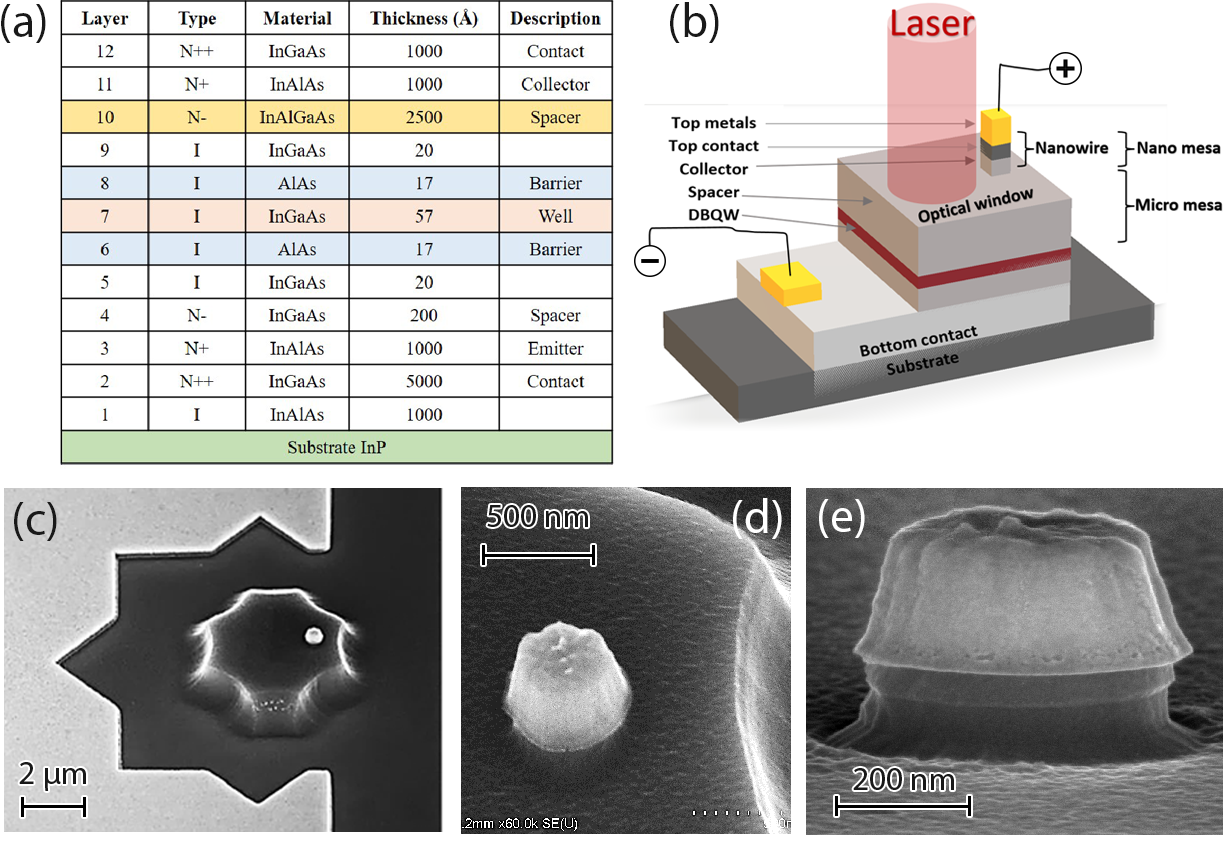}
    \caption{(a) Epitaxial layer structure of the nanostructure RTD-PD highlighting the DBQW structure and the InGaAs photoconductive spacer region. (b) Schematic diagram of the nanostructure RTD-PD with a top collector nanopillar; light is injected into the photoconductive spacer layer vertically. Forward biasing is assumed in this diagram. (c-e) SEM images of the nanostructure RTD-PD, with the bottom metal contact and a detailed view of the top collector nanopillar (d,e).}
    \label{fig:fig1}
\end{figure}

The RTD epitaxial wafer is grown on a semi-insulating InP substrate using molecular beam epitaxy (MBE). Fig. \ref{fig:fig1}(a) presents the detailed layer structure of the wafer. The core of the RTD-PD employs a double barrier quantum well (DBQW) structure comprising a \SI{5.7}{\nano\metre} In$_{0.53}$Ga$_{0.47}$As quantum well sandwiched by two \SI{1.7}{\nano\metre} AlAs barriers. A \SI{250}{\nano\metre} lightly doped InAlGaAs spacer layer was incorporated on the collector side of the DBQW region for light absorption in the infrared wavelength range. The collector and emitter layers are made of \SI{100}{\nano\metre} Si-doped InAlAs layers. The RTD incorporates a nanopillar top contact region with a diameter of $D_{TC} = $ \SI{500}{\nano\metre}, corresponding to top contact area $S_{TC} = $ \SI{0.2}{\micro\metre}$^2$. The structure is visualized on scanning electron microscope (SEM) images in Fig. \ref{fig:fig1}(c-e). To improve the photodetection response, a sufficient illumination area is required. Therefore, the top optical window of the photoconductive spacer layer of the device is purposely kept with a micrometre-scaled size, with total star-shaped device area of $S_{total} =46 \ \mu$m$^2$). 

For comparison, peak (dark) current value for the nanopillar-incorporating device presented in this work (Fig. \ref{fig:IV}) is $I_{P} = $ \SI{11}{\milli\ampere} (see Fig. \ref{fig:IV}), yielding current density (per $S_{TC}$) as $I_{P}/S_{TC}=$\SI{55}{\milli\ampere}/$\mu m^2$ and current per total device area $I_{P}/S_{total}= $ \SI{0.24}{\milli\ampere}/$\mu m^2$. For comparison, in previous work \cite{Lourenco2022_JPCS}, we have designed and fabricated RTD-PD devices built with the same epilayer structure, but with micrometre-scale mesa sizes of $S_{total}= 20\times20$ $\mu$m$^2$ and $S_{total}=$ $10\times10$ $\mu$m$^2$. These larger micrometre-scale RTD-PDs exhibited exhibit $I$-$V$ curves with peak currents of up to \SI{60}{\milli\ampere} for the $10\times10$ $\mu$m$^2$ samples and nearly \SI{250}{\milli\ampere} for the $20\times20$ $\mu$m$^2$ samples respectively. This $10\times10$ $\mu m^2$ RTD device without nanopillar with same spacer size \cite{Lourenco2022_JPCS} exhibited $I_{P}/S_{total} = $ \SI{0.6}{\milli\ampere}/$\mu m^2$. Therefore, the new design is estimated to reduce the total current per device area by $\approx$ $60\%$. 

In addition, since the photoconductive spacer layer remains unchanged with a micrometre-scaled mesa size, the generated photocurrent becomes prominent compared to the dark current, is less affected by thermal noise and can therefore result in the enhancement of optical responsivity. We have also observed comparable improvement in bias stability using the nanopillar mesa. Moreover, further downscaling can lead to RTDs with ultra-low currents \cite{Al-Taai2022_21EMICCE}.

%\hlr{Should we discuss any advantages of using the star-shape for the device?} 
%From Qusay's point of view, the sharp edges of the star shape works as a charges accumulations points and this leads to produce an internal uniform electric field helps to increase the device performance. But all these factors need mor of research to be explained clearly.

\section{Device characterisation under dark and light-illumination conditions}
\begin{figure}[t]
    \centering
    \includegraphics[width=0.99\linewidth]{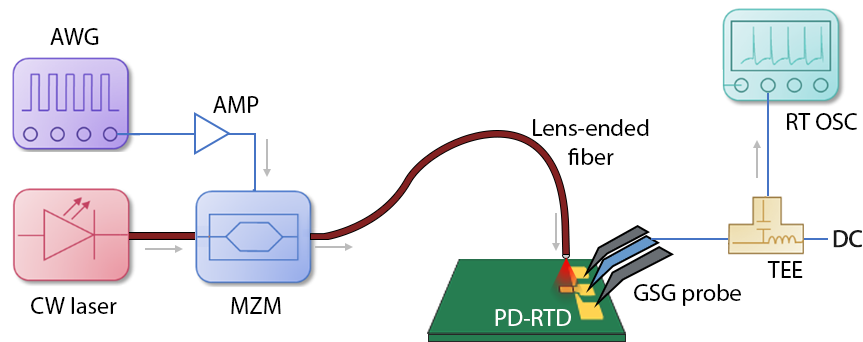}
    \caption{Experimental set-up for photodetection and optical-induced spike generation in a nanostructure RTD-PD.}
    \label{fig:setup}
\end{figure}

First, we characterize the photodetection properties of the fabricated RTD-PD under the injection of an optical signal from a continuous wave (CW) laser at the telecom wavelength of \SI{1310}{\nano\metre}. The experimental setup is depicted in Fig. \ref{fig:setup}. This shows the elements to drive the nanostructure RTD-PD as well as the pathway of the optical input signal. The latter is injected into the device from its top surface using a lens-ended optical fibre (with a spot diameter of \SI{6.5}{\micro\metre}, $\approx$92\% of the size of the device's optical window). An average optical power of approx. \SI{1}{\milli\watt} is used for characterizing the photodetection performance and the dynamical modulation evaluation of the nanostructure RTD-PD in this work. 
For the dynamical analysis of the light-induced spiking regimes in the device, the incoming light from the \SI{1310}{\nano\metre} laser is externally modulated by a Mach-Zehnder modulator (MZM). This encodes the optical input with intensity perturbations generated by an arbitrary waveform generator (AWG). The nanostructure RTD-PD under test has been designed with ground-signal-ground (GSG) bond pads connecting to the emitter (two ground pads) and collector (the signal pad) electrodes (both are \textit{n}-doped) of the device’s epi-layer structure. The RTD-PD is connected to the voltage source via a GSG RF probe landed on the corresponding bond pads. A DC voltage is applied to the device in the reverse biasing direction, which is defined as the RTD collector electrode being linked to the negative pole of the DC source whose voltage is swept from \SI{0}{\volt} up to \SI{800}{\milli\volt}.

% MZM realize optical perturbations with amplitude (as recorded on photodetector) approx. $30\%$ above and $40\%$ below the continuous optical power when using full available range of the AWG digital-to-analog converter (DAC).

\begin{figure}[h]
    \centering
    \includegraphics[width=0.5\linewidth]{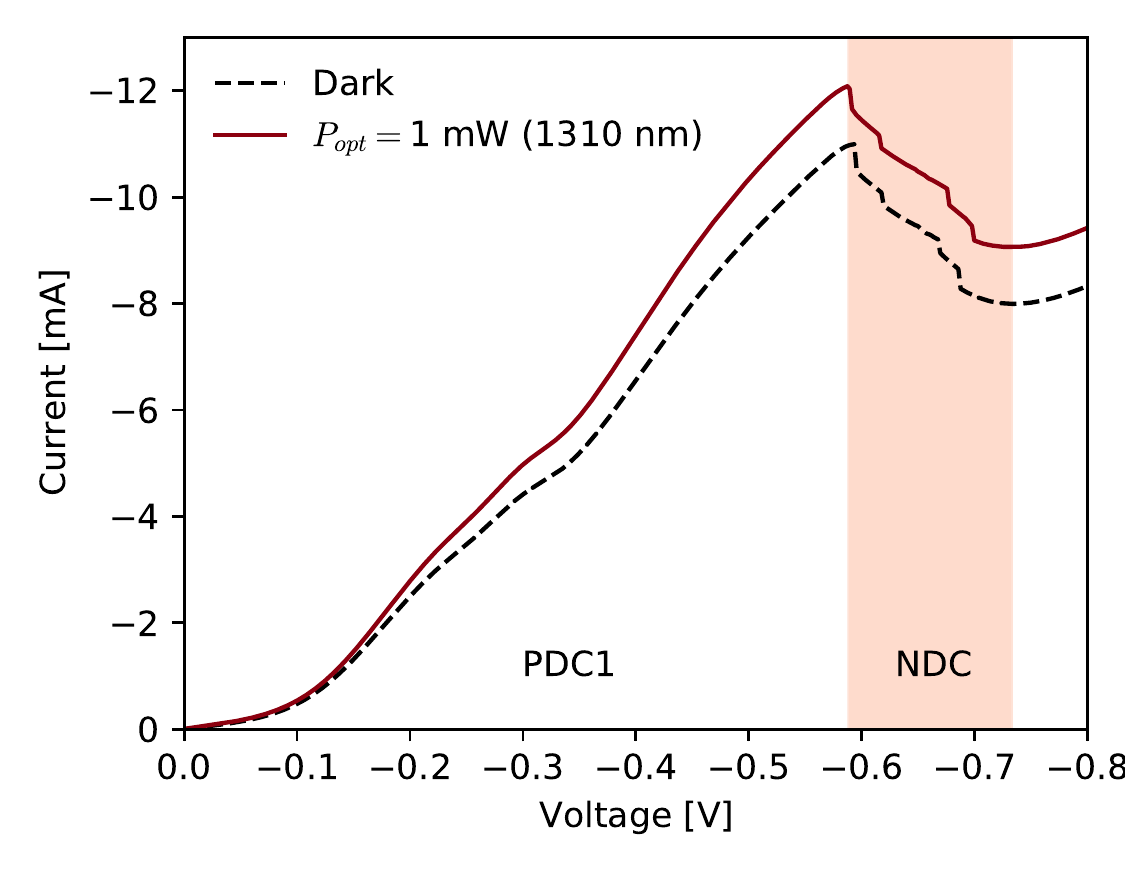}
    \caption{Static $I$-$V$ curves of the nanostructure RTD-PD measured in the dark condition (black, dashed) and with light illumination (red, solid line). The nanostructure RTD-PD is reverse-biased.}
    \label{fig:IV}
\end{figure}

Fig. \ref{fig:IV} illustrates the static $I$-$V$ curve of the nanostructure RTD-PD, measured under dark conditions (without input light signal, blue trace) and when subject to illumination from a \SI{1310}{\nano\metre} continuous wave laser light with average optical power of \SI{1}{\milli\watt} (orange trace). Both $I$-$V$ response curves in Fig. \ref{fig:IV} exhibit the characteristic \textit{N}-shape exhibited by RTDs \cite{Cornescu2019_ITMTT}, with different regions of positive and negative differential conductance (PDC and NDC) defining the two distinct operation points. These are commonly referred to as the 'peak' and 'valley' operation points and are located at the inflexions between the first PDC and the NDC, and the NDC and the second PDC, respectively. Under dark conditions, the $I$-$V$ curve of the device exhibits an NDC region with a peak point of $\approx$\SI{11}{\milli\ampere} and voltage of \SI{593}{\milli\volt}. 

Under illumination, the $I$-$V$ curve typically exhibits both a horizontal (up) and a vertical (towards left) shift. The effect of the $I$-$V$ shift can be attributed to the photogenerated hole accumulation in the collector region in the vicinity of the resonant tunnelling structure (DBQW), as well as the enhanced photoconductivity \cite{Pfenning2016_Nanotechnology, Hartmann2012_APL}. For the tested device under illumination, the peak point under illumination is located at (\SI{587}{\milli\volt} and \SI{12}{\milli\ampere}) with no significant horizontal shift of the $I$-$V$ observed. This corresponds to estimated observed photoresponsivity at \SI{1310}{\nano\metre} as approx. \SI{1.1}{\ampere/\watt}. The demonstrated capability of photodetection and the light-induced $I$-$V$ shifting underpin the functionality of the nanostructure RTD-PD of this work as optically excitable spike generator. The choice of \SI{1}{\milli\watt} was made to observe a strong, well-defined light-induced shift in the device’s $I$-$V$ curve during aligning with the current, proof-of-concept version of the setup (where losses are present due to fibre-coupling, surface reflectivity, etc., which allow for only fraction of the total optical input light to be absorbed in the device). With further optimisation of the experimental setup, operation at sub-mW (and below) optical injection powers can be achieved with current devices. Multiple nanopillar injectors may also be utilized to increase total dark current and allowing for better device performance in terms of responsivity.

% When biasing into the NDC region, the tunneling current decreases and spurious self-oscillations emerge, resulting in a measurable RF output. Beyond the NDC region, the self-oscillations cease and the current of RTD monotonally rises again with increase in applied voltage.

\section{Optically induced excitability in nanostructure RTD-PDs: deterministic spike generation}

\begin{figure}[htb]
    \centering
    \includegraphics[width=0.8\linewidth]{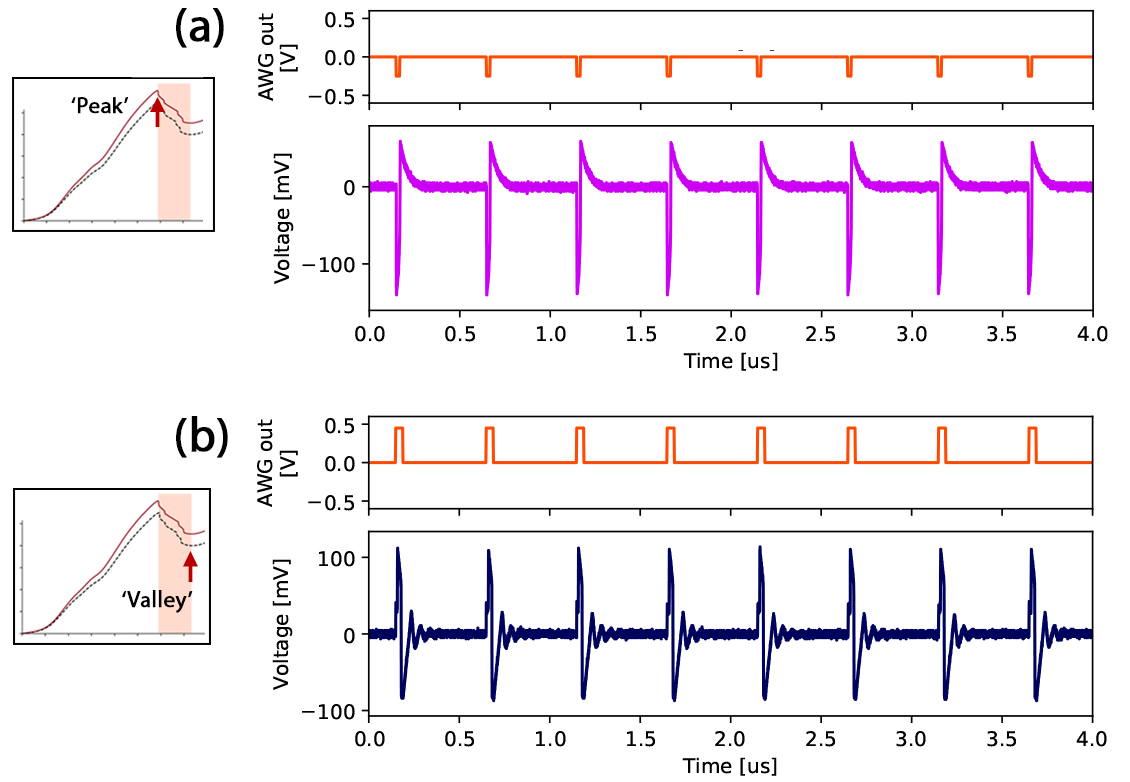}
    \caption{Deterministic, optically-triggered spiking in the nanostructure RTD-PD under peak (a) and valley (b) operation. Time traces in orange (top plots in (a,b)) depict the input optical signals with encoded super-threshold negative (positive) optical perturbations for peak (valley) operation. The bottom plots in (a,b) show the recorded time series at the device’s output when operated in the peak (a, purple) and valley (b, blue). All-or-nothing electrical spikes are fired in response to all of the incoming optical perturbations. The device is reverse biased with a voltage of $V_0=$ \SI{593}{\milli\volt} ($V_0=$ \SI{690}{\milli\volt}) when operated in the peak (valley). Input optical signal is injected with \SI{1}{\milli\watt} avg. optical power (at \SI{1310}{\nano\metre}) with encoded optical perturbations (\SI{15}{\nano\second} long optical pulses, \SI{500}{\nano\second} repetition interval).}
    \label{fig:repeating}
\end{figure}

\begin{figure}[htb]
    \centering
    \includegraphics[width=0.8\linewidth]{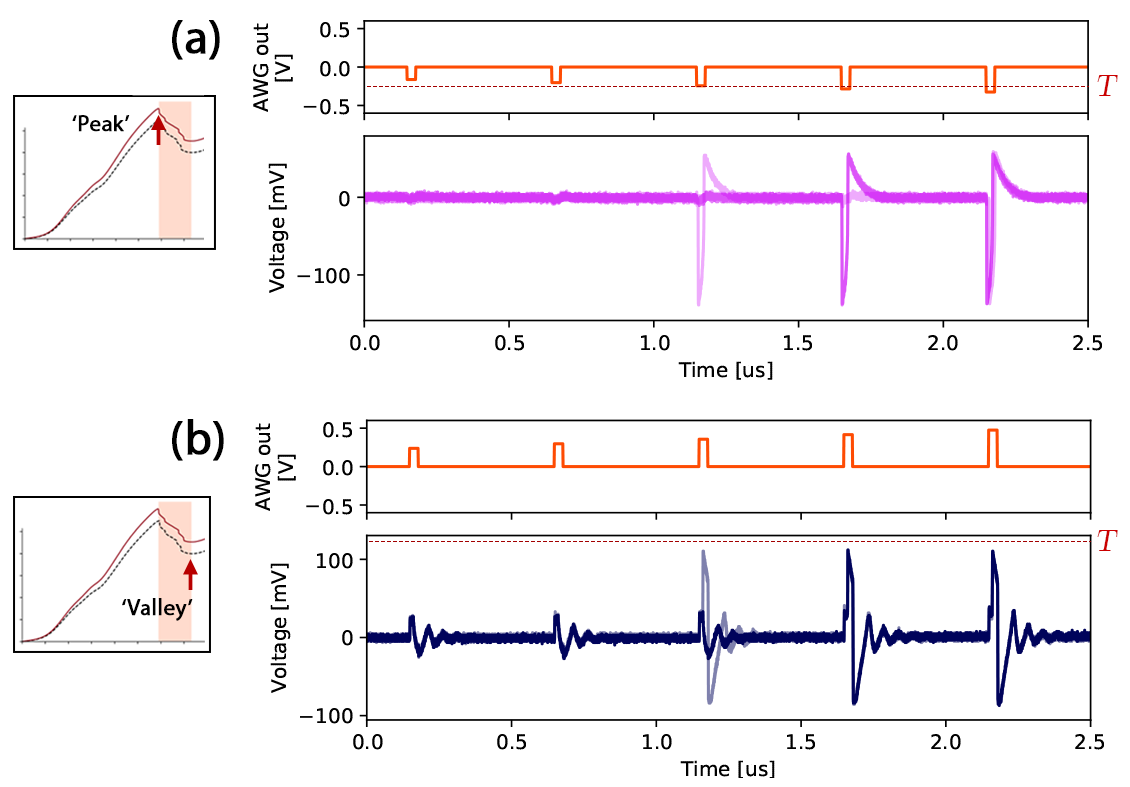}
    \caption{Thresholding of optically-induced spiking in a nanostructure RTD-PD under peak (a) and valley (b) biasing operation. Orange time traces in the top plots in (a,b) depict the input optical signals with encoded negative (positive) optical perturbations of increasing intensity for peak (valley) device operation. The bottom plots in (a,b) depict the recorded time series at the output of the device (10 superimposed traces are given) when operated in the peak (a, purple) and valley (b, blue). The nanostructure RTD-PD fires all-or-nothing electrical spikes only when the input perturbations exceed the excitability threshold (T, dashed orange lines in the top plots in (a,b)). The device is reverse biased with a voltage of $V_0=$ \SI{593}{\milli\volt} ($V_0=$ \SI{690}{\milli\volt}) when operated in the peak (valley). Input optical signal is injected with \SI{1}{\milli\watt} avg. optical power (at \SI{1310}{\nano\metre}) with encoded optical perturbations (\SI{15}{\nano\second} long optical pulses, \SI{500}{\nano\second} repetition interval).}
    \label{fig:thresh}
\end{figure}

\begin{figure}[htb]
    \centering
    \includegraphics[width=0.8\linewidth]{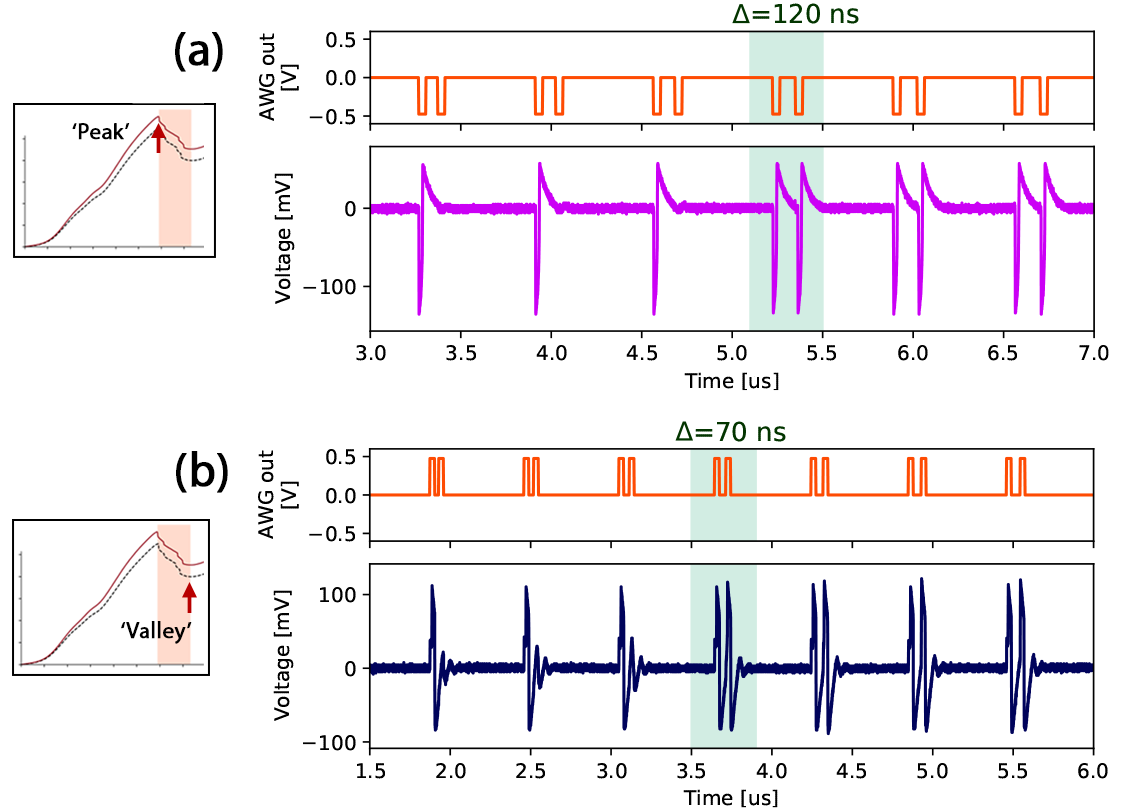}
    \caption{Refractoriness of optically-induced spike firing in the nanostructure RTD-PD when reverse biased in the peak (a) and valley (b) regions. Top plots in (a,b) show (in orange) the input optical signals encoded with optical perturbations in the form of pulse doublets with increasing temporal perturbations between the individual pulses. Bottom plots in (a,b) show measured time traces at the device’s output when operated in the peak (a) and valley (b) points. The device was reverse biased with $V_0=$ \SI{593}{\milli\volt} and $V_0=$ \SI{690}{\milli\volt}, when operated in the peak and valley respectively. Based on the response of the RTD-PD, the refractory period is estimated as $T_{R_P} \approx $ \SI{120}{\nano\second} in the peak, and $T_{R_V} \approx $ \SI{70}{\nano\second} in the valley.}
    \label{fig:refractory}
\end{figure}

To operate the nanostructure RTD-PD as an optically excitable spiking neuron, the device is reverse biased at either the peak or valley operation regions, in close proximity to the NDC boundary. 
In our experiments, we reverse biased the nanostructure RTD-PD with voltage values of $V_0 =$ \SI{593}{\milli\volt}, $V_0 =$ \SI{690}{\milli\volt} for operating in the ‘peak’ or ‘valley’ operation point respectively. 
In both cases, the device is directly subject to the injection of optical input signal (stimuli) in the form of square-shaped optical pulses. The incoming optical perturbations briefly shift the $I$-$V$ curve of the device (see Fig. \ref{fig:IV}), which in turn momentarily shifts the device’s operation point (in both peak or valley biasing cases) into the NDC region. As a result of this brief transition induced by the incoming optical perturbation, an excitable response is elicited and the device fires an electrical spike. This effect underpins the mechanism whereby the nanostructure RTD-PD of this work can behave as an excitable spike generator (acting in fact as an optoelectronic spiking neuron). Importantly, the incoming optical stimuli are referred to as ‘super-threshold’ (or sub-threshold), when their optical intensity exceeds (or does not exceed) a given threshold power required for excitability in the nanostructure RTD-PD device. To obtain the desirable spike firing behaviour in both cases of operation (peak and valley regions), the transition edge of the output electrical spikes is matched to that of the incoming pulse-shaped optical stimuli. Therefore, positive input optical pulses are employed when setting the bias voltage of the nanostructure RTD-PD at the ‘valley’ region, while negative optical pulses (power drops) are used when the device is biased at the ‘peak’. Multiple optical stimuli variations are investigated (see Figs. \ref{fig:repeating} to \ref{fig:refractory}) to demonstrate and experimentally evaluate the performance of the nanostructure RTD-PD as an excitable spike generator. 

Fig. \ref{fig:repeating} shows experimentally captured time series demonstrating the deterministic, controllable and repeatable firing of excitable (electrical) spikes in the nanostructure RTD-PD in response to input optical perturbations. 
Specifically, Figs. \ref{fig:repeating}(a) and \ref{fig:repeating}(b) respectively plot the cases when the device is biased at the peak and valley regions. To illustrate the deterministic spike firing behaviour, the system was tested under the injection of a sequence of consecutive optical perturbations, shown in the top plots (orange time traces) in Figs. \ref{fig:repeating}(a) and \ref{fig:repeating}(b). These were configured as optical input rectangular-shaped pulses (with positive/negative amplitudes for operation in the peak/valley) with a temporal length of \SI{15}{\nano\second} and a repetition interval of \SI{500}{\nano\second}. The bottom plots in Figs. \ref{fig:repeating}(a) and \ref{fig:repeating}(b) show the recorded electrical time series measured at the output of the nanostructure RTD-PD, when this was operated in the peak (Fig. \ref{fig:repeating}(a)) and valley (Fig. \ref{fig:repeating}(b)) regions, respectively. These demonstrate for both cases of operation that the nanostructure RTD-PD produces deterministic and repeatable all-or-nothing spike events, all with the same shape and temporal features, in response to every single one of the incoming optical perturbations. Figs. \ref{fig:repeating}(a) and \ref{fig:repeating}(b) also show that as expected, the device fires spikes of opposite polarity for both cases of operation (peak or valley). In the case of valley operation, slightly more pronounced relaxation oscillations can be observed. We believe that these can be mainly attributed to the more shallow shape of the $I$-$V$ in the valley region.
%Shown in previous works: 

Fig. \ref{fig:thresh} plots measured time traces at the output of the nanostructure RTD-PD when biased in the peak (Fig. \ref{fig:thresh}(a)) and valley (Fig. \ref{fig:thresh}(b)) regions and under the injection of optical input perturbations of gradually increasing intensity. 

Specifically, the bottom plots in Figs. \ref{fig:thresh}(a) and \ref{fig:thresh}(b) each include an overlay of $n=10$ recorded output time traces, when the device is reverse biased in the peak and the valley points, respectively. The input optical signal (at the wavelength of \SI{1310}{\nano\metre} with \SI{1}{\milli\watt} of average optical power) is modulated with a sequence of perturbations (\SI{15}{\nano\second} long optical pulses at \SI{500}{\nano\second} repetition period). The amplitude of these input optical pulses (negative or positive, for peak or valley device operation) increases monotonically, until a threshold is reached upon which the nanostructure RTD-PD elicits all-or-nothing excitable electrical spikes. Incoming perturbations below the threshold level (sub-threshold) are not able to elicit spike firing and the system therefore remains quiescent. We must note here that when the strength of the incoming optical brings the system just at the spike firing threshold level, stochastic spike firing may occur. Such is the case for the system’s response to the third input optical perturbation in Figs. \ref{fig:thresh}(a) and \ref{fig:thresh}(b). In that specific case, as the incoming perturbations bring the system just at the excitability threshold, spike events are achieved with probability below unity (both for peak and valley operation). Hence, for that case Figs. \ref{fig:thresh}(a) and \ref{fig:thresh}(b) show that spikes are only elicited for some of the $n = 10$ superimposed measured output traces. However, as soon as the optical inputs exceed the excitability threshold (4th and 5th optical perturbations) consistent and deterministic spike events are elicited by the nanostructure RTD-PD element. Again, the shape and temporal features of the elicited electrical spikes are independent of the amplitude of the incoming optical perturbations. We must also note that the spike thresholding level (for both peak and valley operation) can be tuned at will simply by acting on the voltage applied to the nanostructure RTD-PD, to either bias it closer or further from the NDC transition. Hence, the strength of the optical input stimuli required to elicit spike firing events would be lower if the bias voltage is set closer to the boundary of the NDC region, and vice versa. This offers a simple, yet powerful mechanism to control the system’s spiking response. 

% Notably, the strength of the stimuli (optical intensity) required for reaching the excitability threshold depends on the bias voltage, which is fixed by the DC bias voltage. Relatively lower optical power is required if the bias voltage is set closer to the boundary of the NDC region, and vice versa. Furthermore, it is worth noting that when the strength of the incoming stimuli is very close above the sub-threshold/super-threshold boundary, the spiking process will be stochastic as the RTD-PD may not fire spikes with certainty, such as the 3rd spike in Fig. \ref{fig:thresh}(a) and Fig. \ref{fig:thresh}(b). The difference in used pulse voltage amplitudes in the experiments ($\approx$\SI{300}{\milli\volt}) amplitude to reach threshold in the peak, $\approx$\SI{400}{\milli\volt} to reach threshold in the valley) comes primarily from the set operation point of the MZM (between quadrature and peak power), in which power drops are elicited with lower voltage signals than comparable upward pulses.

Fig. \ref{fig:refractory} provides the experimental evaluation of the refractory period of the spiking responses in the nanostructure RTD-PD. Again, results are given for the cases of system operation in the peak (Fig. \ref{fig:refractory}(a)) and valley (Fig. \ref{fig:refractory}(b)) regions. If an excitable system (such as a biological neuron) fires a spike event in response to an incoming perturbation, it does not fully recover immediately. Instead, it takes a period of time (known as the refractory period), in which it cannot elicit new spiking events even if new super-threshold perturbations enter the system \cite{Hejda2022_Nanophot}. Hence, this refractory period practically governs the maximum achievable spike firing rate of an excitable system. Fig. \ref{fig:refractory} evaluates whether this neuron-like dynamical property, refractoriness, is also observed in the nanostructure RTD-PD of this work. To do so, input optical perturbations in the form of super-threshold square optical pulse doublets are injected into the device. The top plots (orange traces) in Figs. \ref{fig:refractory}(a) and \ref{fig:refractory}(b) show that individual pulses in the doublets are provided with gradually increasing temporal separation. 
In the case of device operated in the peak region (Fig. \ref{fig:refractory}(a)), the doublets consist of two individual \SI{40}{\nano\second} long pulses, and their temporal separation (measured between pulse rising edges) is gradually increased from \SI{98}{\nano\second} to \SI{134}{\nano\second} in steps of \SI{7}{\nano\second}. 
The purple trace in the bottom plot of Fig. \ref{fig:refractory}(a) shows the recorded output from the nanostructured RTD-PD when biased in the peak region with a reverse bias voltage of $V_0=$ \SI{593}{\milli\volt}. This shows that at first when the optical pulses in the doublet are very close to each other, the system only fires a spike in response to the first optical input pulse. The second input pulse arrives within the refractory period of the system and therefore a second spike is not produced. As the temporal separation between the individual pulses in the doublet is increased, a clear transition can be observed and the system’s output changes from a single spike to a double spike firing pattern. This transition demonstrates the existence of a refractory period in the system (as in biological neurons) below which it cannot elicit a second spike after firing a first spike event. Based on the experimental findings in Fig. \ref{fig:refractory}(a), this refractory period for the nanostructure RTD-PD when operated in the peak is estimated as $T_{R_P} \approx $ \SI{120}{\nano\second}. The same dynamical behaviour is also observed when the device is operated in the valley region (see Fig. \ref{fig:refractory}(b)). In this case, optical input pulse doublets formed by \SI{30}{\nano\second} long pulses with gradually increasing temporal separation from \SI{55}{\nano\second} to \SI{85}{\nano\second} in increments of \SI{5}{\nano\second}. A clear refractory period is also observed in the valley operation case, with the system transitioning from a single spike to a two-spike pattern output when the temporal separation between pulses exceeds approx. $T_{R_V}\approx$ \SI{70}{\nano\second}. A shorter refractory period is observed when the system is operated in the valley than when it is operated in the peak. This difference in the measured refractory periods in both cases is believed to be due to circuit parameters (such as capacitance and series resistance \cite{Ironside2019_RTDbook}) being dependent on the applied bias. Furthermore, the asymmetric character of the NDC region (Fig. \ref{fig:IV}) in the nanostructure RTD-PD also contributes to variation in the character of dynamical behaviours between the two operation points.

These experimentally measured refractory periods are in the order of \SI{100}{\nano\second}, defining readily demonstrated maximum firing rates of approx. \SI{10}{\mega\hertz} in our system. Whilst being already orders of magnitude faster than those of biological neurons and also significantly faster than previously reported in a similar, non-nanostructure system \cite{Zhang2021_Nanomaterials}, these firing rates are by no means the maximum achievable speeds expected for this new nanostructure device class. The refractory time is limited here by parasitic effects arising from the external electrical circuitry (e.g. cables, bias-tees, etc.) used in our laboratory setup to characterise the system’s light-induced spiking functionality. However, the resonant tunnelling effect underpinning the operation of RTD systems is an ultrafast phenomena shown to enable RTDs oscillators at up to THz rates at room temperatures \cite{Asada2016_JIMTW}. For photodetecting RTD, carrier dynamics will likely impose limitations to bandwidth of optical response to few tens of GHz, which still presents potential for significant increase in operation speeds of these nanopillar RTD-PD devices.

Moreover, recent numerical reports from our group \cite{Hejda2022_Nanophot} have shown that nanostructure RTD-PDs are able to achieve (light-induced) high-speed spiking rates down to sub-ns regimes, with numerically predicted refractory times as low as \SI{300}{\pico\second}, hence enabling multi-GHz spiking rates. Therefore, future work aiming at developing optimised nanostructure RTD-PD elements with optimized, high-speed circuitry is expected to dramatically improve the speed operation for light-induced spiking regimes in nanostructure RTD-PD systems, beyond the current fast (approx. \SI{10}{\mega\hertz}) spiking rates demonstrated already in this first proof-of-concept demonstration. Similarly, whilst this work demonstrates light-induced spiking generation using optical powers of approx. \SI{1}{\milli\watt}, further optimisation stages to reduce coupling losses in the setup as well as to improve light absorption in the device are expected to yield operation at significantly lower optical injection powers, well below the current \SI{1}{\milli\watt} level. This added to the reduced bias voltages and current enabled by the nanopillar-based design of this new nanostructure device class, will permit the realisation of high-speed (>GHz rates) optoelectronic spiking neurons for novel light-enabled neuromorphic computing hardware.  

\section{Conclusions}
 
 This work reports for the first time deterministic, optically-triggered generation of neural-like excitable spiking signals in a nanostructure resonant tunnelling diode with integrated photo-detecting capability (RTD-PD). This is formed by a single \SI{500}{\nano\metre} diameter nanopillar on top conductor structure of the RTD, resulting in a reduced peak operation point current of only \SI{11}{\milli\ampere}. This is about $60\%$ lower than comparable devices without a nanopillar structure, allowing for more efficient operation as a spike generator. The embedded light absorptive spacer layer adjacent to the DBQW structure at the device core endows the device with high-responsivity photodetection, therefore enabling light-induced excitability. The static characterisation of the device under dark conditions and when subject to infrared light illumination reveals a light-induced shift in the device’s region of negative differential conductance (NDC). This key feature underpins the optical triggering of excitable regimes and allows operation of the device as an optoelectronic spiking neuron. Importantly, this new system exhibits a wide range of neural-like behaviours. Among these, we have experimentally demonstrated deterministic optical activation of all-or-nothing spike events at ns-rates, both when the device is operated at the so-called peak and valley points of its IV curve (close to the NDC region boundaries). Furthermore, as in biological neurons, we have also shown that the optically-triggered spiking signals from the nanostructure RTD-PD exhibit a clear threshold for spike firing, as well as well-defined refractoriness. Our experimental findings reveal already in this first proof-of-concept demonstration refractory periods in the order of \SI{100}{\nano\second}, hence enabling firing rates of $\approx$\SI{10}{\mega\hertz}; thus, already multiple orders of magnitude faster than in biological neurons. Current operation speed is mainly imposed by limitations of the external circuitry in our current setup, and it is by no means the limit in the expected spike firing rates in this system. Resonant tunnelling is an ultrafast phenomena with reported RTD bandwidth exceeding one THz at room temperature. Hence, as documented in previous numerical works \cite{Hejda2022_Nanophot}, optically-triggered firing rates at GHz rates (and possibly higher) should be possible in the nanostructure RTD-PD with future device and system optimisation stages. These exciting neuromorphic features, added to their light-enabled capabilities, render this nanostructure RTD-PD as a promising new device class for the realisation of novel fast, energy-efficient optoelectronic neurons for light-enabled neuromorphic computing architectures.

\section{Acknowledgement}
The authors acknowledge support from the European Commission (Grant 828841-ChipAIH2020-FETOPEN-2018-2020) and by the UK Research and Innovation (UKRI) Turing AI Acceleration Fellowships Program 4-(EP/V025198/1).

%%%%%%%%%%%%%%%%%%%%%%% References %%%%%%%%%%%%%%%%%%%%%%%%%

\bibliography{main}

\end{document}